\documentclass{aastex631}

\begin{document}

\title{A New Look at the YY CrB Binary System}

\author{Somayeh Soomandar}
\affiliation{Independent astrophysics researcher, Kerman,
Iran
}

\author{Atila Poro}
\affiliation{Astronomy Department of the Raderon Lab., BC., Burnaby, Canada
}

\begin{abstract}
This study presented a new analysis for the TESS-observed W Ursae Majoris (W UMa) binary star YY Coronea Borealis (YY CrB). The light curve was analyzed by the PHysics Of Eclipsing BinariEs (PHOEBE) Python version together with the Markov chain Monte Carlo (MCMC) method. The light curve solutions required a hot spot and $l_3$. New eclipse times from the TESS observations were extracted, and the O-C curve of primary and secondary minima showed an anti-correlated manner. In order to study the O-C curve of minima, minima times between 1991 and 2023 were collected. This investigation reported a new linear ephemeris and by fitting a quadratic function to the O-C curve of minima, calculated the orbital period rate of $\mathop P\limits^.  \approx 5.786\times{10^{-8}}\frac{{day}}{{year}}$. Assuming mass conservation, a mass exchange rate of $\mathop{{M_2}}\limits^.=2.472\times{10^{-8}}$ calculated from the more massive component to the less massive one. Then, by using the light travel time function, the possible third body was determined in the binary and derived the mass of the third body as $0.498{M_\odot}$ with a period of $\simeq 7351.018$ days. The O-C curve analysis and the quantity of mass indicate that the presence of a third body is unlikely. This binary is expected to evolve into a broken-contact phase and is a good case to support the thermal relaxation oscillation model.
\end{abstract}


\vspace{1.5cm}
\section{Introduction} \label{sec:intro}

W UMa-type systems are recognised by their eclipsing light curves with almost equal minima and a short orbital period. These stars have spectral types ranging from A to middle K, and the convective atmosphere is the main reason for chromosphere activity, as well as starspots, which are signs of the existence of dynamo-generated magnetic activity.

YY CrB (HIP 77598, TIC 29287800) is a W UMa binary system discovered by Hipparcos \citealt{1997ESASP1200.....E}. This system has been studied in some works; first \citet{2000AJ....120.1133R} revealed the spectral type F8V for two components.\citet{2004BaltA..13..151V} found the light curves to be asymmetric and mentioned the existence of starspots on the components. \citet{2005AcA....55..123G} analyzed the light curve and derived the geometric and photometric parameters and concluded that this target is a contact binary with weak magnetic activity.

\citet{2010NewA...15..227E} combined photometric and spectroscopic solutions and calculated the fill-out factor approximately equal to 64 percent and mass ratio of 0.241. In addition, they studied the changes in the orbital period using the O-C diagram and concluded that the orbital period is decreasing. \citet{2015PASJ...67...42Y} studied the orbital period changes and implied that the decreased period rate and concluded that the sinusoidal oscillatory can be interrupted as magnetic activity. \citet{2010NewA...15..227E} and \citet{2015PASJ...67...42Y} on the rate of period decrease demonstrate that the value of reduced rate is lowering progressively, indicating that this system was going through an orbital expansion stage of thermal relaxation oscillation (TRO) cycles.
Also, understanding the evolutionary status of this target could prove invaluable.
Using new space-based data, we have re-analyzed the light curve solution and studied the O-C curve in detail. Moreover, we studied the possibility of a third body in this interesting system.
The structure of the paper is as follows: Section 2 provides information on TESS observations and a data reduction process. The light curve solution and the estimation of absolute parameters are included in Sections 3 and 4 respectively, the orbital period variation analysis is presented in Section 5, and finally, Section 6 contains the discussion and conclusion.
\vspace{1.5cm}
\section{Observation and Data Reduction} \label{sec:style}
\begin{figure*}
  \includegraphics[scale=0.40]{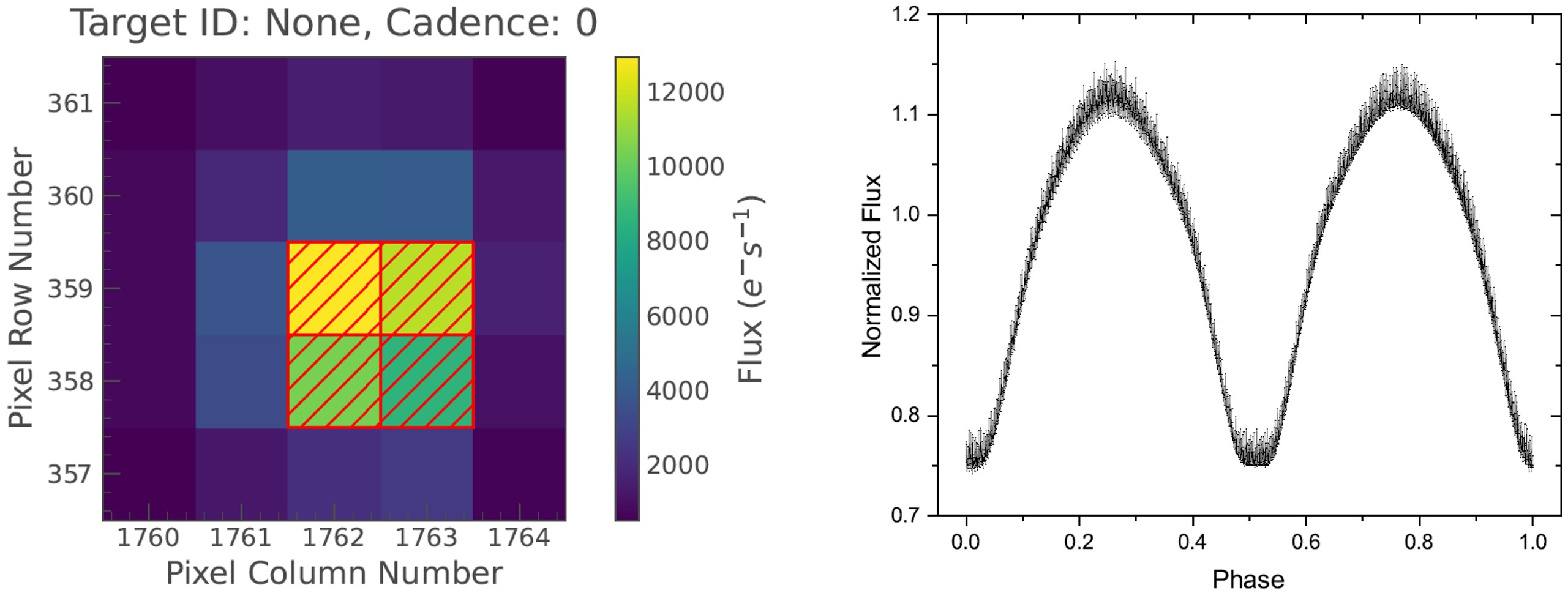}
  \caption{Left panel: TESS target pixel file of YY CrB in sector 51. The pixels included in the computation of the SAP as red bordered pixels. Right panel: phased light curve during sector 51.}
  \label{Fig1}
\end{figure*}

YY CrB was observed by the TESS during sectors 24 and 51 (April 16, 2020-May 13, 2020, and April 22, 2022-May 18, 2022) on Cameras 1 and 3. There is two-minute cadence data for sector 24 that are processed by the Science Processing Operations Center (SPOC) pipeline (\citet{2016SPIE.9913E..3EJ, 2015ESS.....310605J}). Photometric photos were downloaded using the Lightkurve package \citep{2018ascl.soft12013L} that provides the functions to download TESS data from the public data archive at MAST\footnote{\url{https://mast.stsci.edu}}. For sector 24, we used the Pre-search Data Conditioning flux of the Simple Aperture Photometry (PDCSAP). There is no detrended light curve for sector 51. Therefore, we download the TESS Full Frame Images (FFIs) from the MAST and used Lightkurve package to extract the SAP light curve with a mask that are defined by the pixels shown in the left panel of Figure \ref{Fig1}. We used \texttt{create threshold mask} function to produce an aperture mask using a threshold equal to 10. This function identifies the pixels in the target pixel file and shows that a median flux that is brighter than the threshold times the standard deviation above the overall median. The right panel of Figure \ref{Fig1} shows the phased light curve that was produced.
\vspace{1.5cm}
\section{Light curve solution}
\label{sec:command}
\citet{2010NewA...15..227E} calculated the optimal parameters by combining simultaneous radial velocity and light curve solutions. We started with the initial values of the parameters taken from the solution by the \citet{2010NewA...15..227E} study. One-day data from TESS sector 24 were utilized to light curve solution. The observation of 2-min cadence help to better analysis of the effect of spots on the components. Photometric analysis of the YY CrB system was carried out using  the PHOEBE 2.4.9 version, TESS filter of the code, and the MCMC approach (\citealt{2005ApJ...628..426P}, \citealt{2016ApJS..227...29P}, \citealt{2020ApJS..250...34C}, \citealt{2022PASP..134f4201P}).

We selected the TESS passband from the code and chose the contact binary mode in the PHOEBE based on the light curve's shape and solutions of previous studies.

The initial and input parameters were as follows: The mass ratio $q=0.241$ and the effective temperature of primary component ${T_1}=5819$ are \citep{2010NewA...15..227E}, the gravity darkening coefficients, ${g_1}={g_2}=0.32$ and the albedo coefficients, ${A_1}={A_2}=0.5$ (\citealt{1967ZA.....65...89L}, \citealt{1969AcA....19..245R}).

\begin{figure}
\begin{center}
  \includegraphics[scale=0.60]{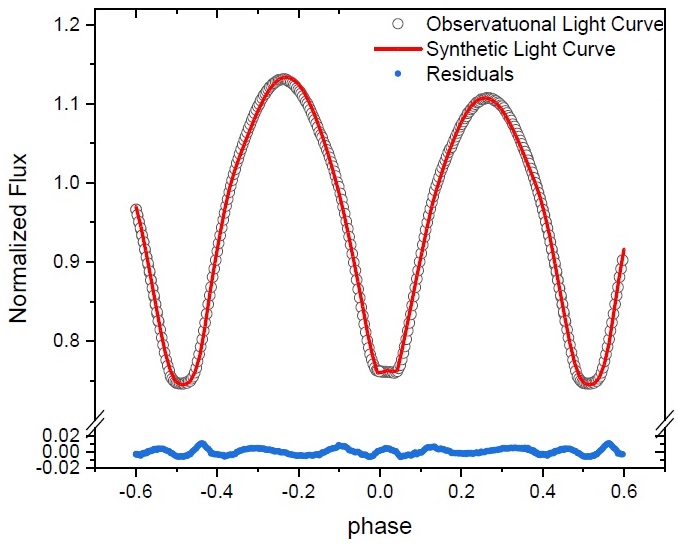}
  \caption{Light curve solution of the eclipsing binary YY CrB. Observational light curve (blank circle), synthetic light curve (solid red line), and the residuals (blue circle).}
  \label{Fig2}
    \end{center}
\end{figure}

\begin{figure*}
\begin{center}
  \includegraphics[scale=0.54]{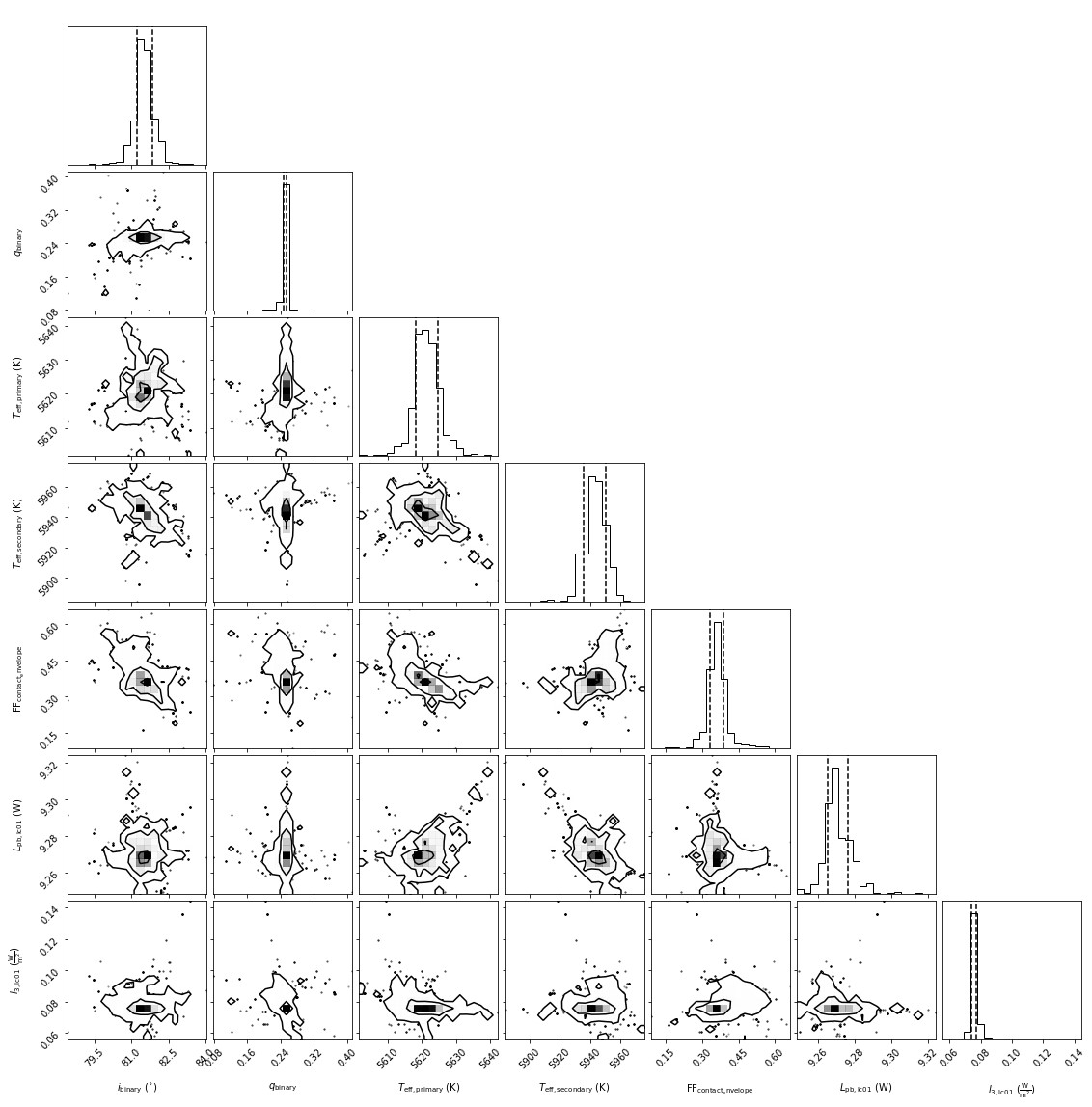}
  \caption{The corner plots of the light curve solution.}
  \label{Fig3}
  \end{center}
\end{figure*}

\begin{figure*}
\begin{center}
  \includegraphics[scale=0.44]{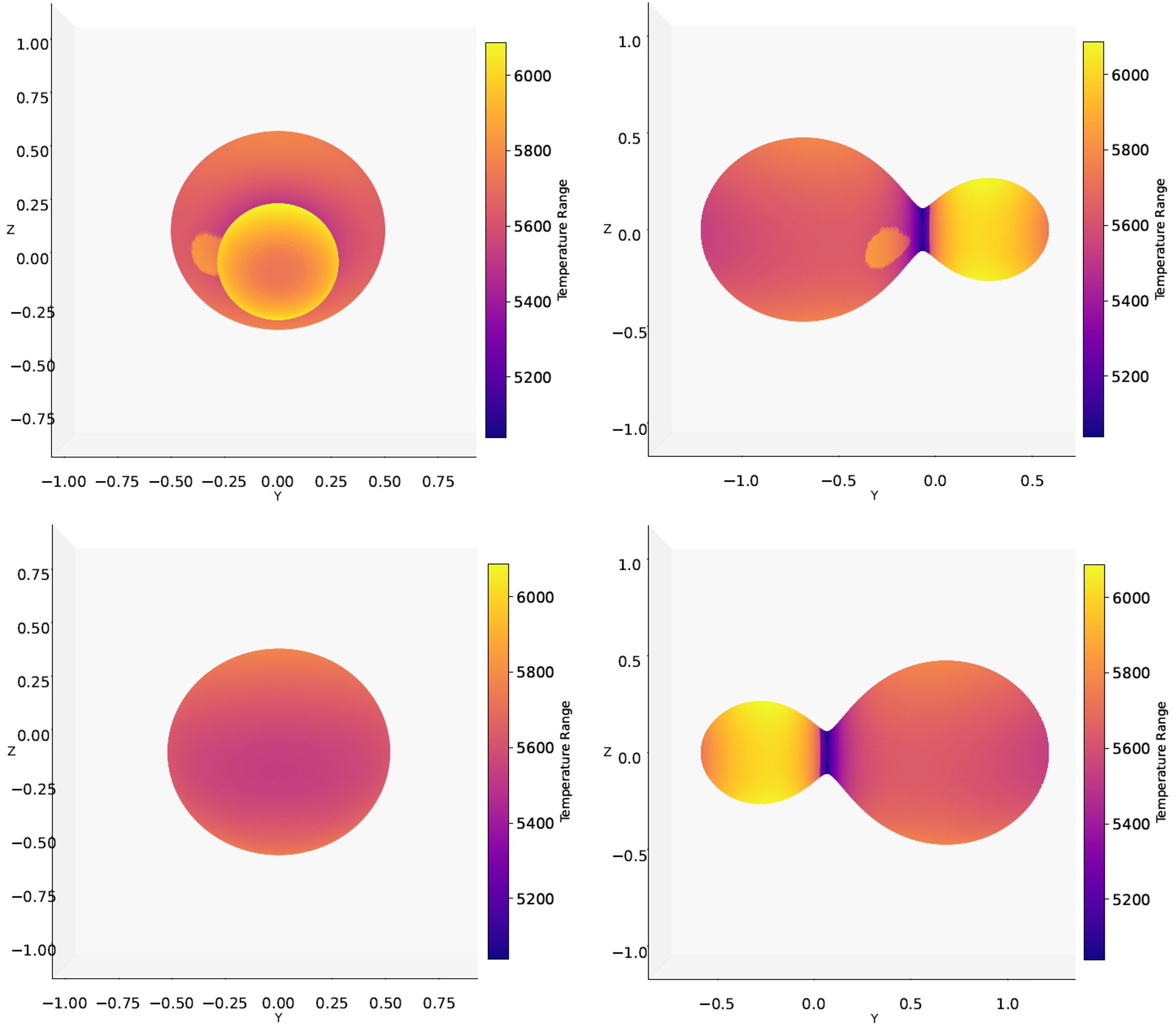}
  \caption{The geometrical structure of YY CrB.}
  \label{Fig4}
  \end{center}
\end{figure*}

The limb-darkening coefficients were employed as free parameters, and the \citet{2004A&A...419..725C} method was used to model the stellar atmosphere. The parameters searched in MCMC include: the orbital inclination $i$, the mean temperature of the stars $T_{1,2}$, the mass ratio $q$, the fillout factor $f$, the bandpass luminosity of the primary star (${L_1}$), and the third light in total light (${l_3}$). We applied 46 walkers and 1000 iterations to each walker in MCMC. According to the asymmetry in the brightness of maxima in the light curve of the close eclipsing binary, the solution requires the assumption of a hot spot on the primary component (\citealt{1951PRCO....2...85O}). According to observational and theoretical light curves in this study, it has not been possible to provide the solution without considering $l_3$. The theoretical fit on the observational light curve for the YY CrB system is given in Figure \ref{Fig2}. The corner plot that MCMC produced is displayed in Figure \ref{Fig3}. Also, the geometrical structure is plotted in Figure \ref{Fig4}, which has a lower temperature at the point of contact between companion stars due to the gravity darkening (\citealt{2016ApJS..227...29P}). The calculated parameters together with the values obtained by \citet{2010NewA...15..227E} are listed in Table \ref{Tab1}.

\begin{table*}\centering
  \newcommand{\DS}{\hspace{6\tabcolsep}}
  \caption{The parameters of the eclipsing binary YY CrB.} \label{Tab1}
  \begin{tabular}{l @{\DS} cccc l cccc}
    \hline
    \hline
    Parameter& This study & \citet{2010NewA...15..227E} \\
    \hline
    $q=M_2/M_1$  & $0.2498_{\rm-0.0024}^{+0.0031}$ & $0.241\pm0.002$\\
    \\
    $T_1$ (K) & $5621_{\rm-3}^{+3}$ & 5819\\
    \\
    $T_2$ (K) & $5944_{\rm-8}^{+6}$ & $6010\pm72$\\
    \\
    $i$ (deg) &  $81.50_{\rm-0.29}^{+0.36}$ & $80.26\pm0.05$  \\
    \\
    ${\Omega_1=\Omega_2}$ &  $2.295\pm0.079$ & 2.237\\
    \\
    $l_1/l_{tot}$ & $0.730_{\rm-0.001}^{+0.001}$ & $ 0.7508\pm0.0154$\\
    \\
    $l_2/l_{tot}$ & $0.264\pm0.001$ & 0.2492\\
    \\
    $l_3/l_{tot}$ & $0.006_{\rm-0.001}^{+0.001}$ & \\
    \\
    $f$ & $0.363_{\rm-0.031}^{+0.025}$ & 0.64\\
    \\
    ${r_1}_{mean}$ & $0.522\pm0.018$ & 0.537\\
    \\
    ${r_2}_{mean}$ & $0.287\pm0.028$ & 0.282\\
    \\
    Phase shift& $0.08\pm0.005$ & \\
    \hline
    Spot on the star 1:\\
         Colatitude $\theta$ & $99\pm1$ & 90\\
         Longitude $\lambda$ & $325\pm1$ & $11.25\pm1.638$\\
         Angular radii $\gamma$ & $18\pm1$ & $5.250\pm0.573$\\
         ${T_{star}}/{T_{spot}}$ & $1.04\pm0.02$ & 0.750\\
         \\
    Spot on the star 2:\\
         Colatitude $\theta$ &  &  $99.487\pm3.919$\\
         Longitude $\lambda$ &  & 325\\
         Angular radii $\gamma$ &  & $16.300\pm1.326$\\
         ${T_{star}}/{T_{spot}}$ &  &  1.351\\
    \hline
    \hline
    \end{tabular}
\end{table*}
\vspace{1.5cm}
\section{Absolute parameters}
The absolute parameters of the binary system including ${M_{v1,2}}$, ${M_{bol1,2}}$, $L_{1,2}$, $R_{1,2}$, $M_{1,2}$, $log(g)_{1,2}$, and $a$ were calculated. We used Gaia DR3 parallaxes and the parameters of the light curve solution in this study. We followed the same method as done by \citet{2022MNRAS.510.5315P}. First the absolute magnitude ${M_v}$ of the system was calculated by Equation ~(\ref{eq:one}).

\begin{equation}
  \label{eq:one}
  {M_{v(system)}} = V_{max} - 5\log (d) + 5 - {A_v}
\end{equation}

where the distance of the system from Gaia DR3 ($d_{pc}=90.07\pm0.1$) was derived and $V_{max}=8.64\pm0.08$ comes from the VSX\footnote{\url{https://www.aavso.org/vsx/}} database. Extinction coefficient ${A_v}=0.015\pm0.002$ was obtained using the DUST-MAPS package in Python \citep{2019ApJ...887...93G}. Also, Equation ~(\ref{eq:two}) can be utilized to determine the primary and secondary components' absolute magnitude.

\begin{equation}
  \label{eq:two}
  {M_{v1,2}} - {M_{vtot}} =  - 2.5\log (\frac{{{l_{1,2}}}}{{{l_{tot}}}})  
\end{equation}

The bolometric magnitude ${M_{bol}}$ of each component of the binary was obtained by Equation ~(\ref{eq:three}),

\begin{equation}
  \label{eq:three}
  {M_{bol}} = {M_v} + BC
\end{equation}

where the effective temperature of the stars is employed to obtain the bolometric correction for the primary and secondary components retrieved $BC_1=-0.111$ and $BC_2=-0.052$ respectively \citep{1996ApJ...469..355F}. The bolometric correction is presented as polynomial fits in Equation ~(\ref{eq:four}).

\begin{equation}
  \label{eq:four}
  BC = a + b(\log {T_{eff}}) + c{(\log {T_{eff}})^2} + d{(\log {T_{eff}})^3} + e{(\log {T_{eff}})^4}
  \end{equation}

Then, the luminosity of two components is determined from Pogson's relation \citep{1856MNRAS..17...12P},

\begin{equation}
  \label{eq:five}
{M_{bol}} - {M_{bol \odot }} =  - 2.5\log (\frac{L}{{{L_ \odot }}})                                                  \end{equation}

where ${M_{bol\odot}}$is taken as $4.73^{mag}$ from \citet{2010AJ....140.1158T}.
The radius of primary and secondary components is calculated by the equation ~(\ref{eq:six}),

\begin{equation}
  \label{eq:six}
R = {(\frac{L}{{4\pi \sigma {T^4}}})^{1/2}}
\end{equation}

where the $\sigma$ is the Stephen-Boltzmann constant and $T$ is the temperatures of each components.

Additionally, with considering ${r_{mean1,2}}$ and $a = \frac{R}{{{r_{mean}}}}$, we calculated the separation $a$ in average ${a_1}$ and ${a_2}$. The resulting parameters and the values obtained by the \citet{2010NewA...15..227E} study are listed in Table \ref{tab2}.

\begin{table}
\begin{center}
\caption{The absolute parameters of YY CrB.}
\centering
\footnotesize
\begin{tabular}{c c c}
 \hline
 \hline
Absolute parameters & This study & \citet{2010NewA...15..227E}  \\
\hline
${M_{bol1}}(mag)$ & $4.083\pm0.074$ & 3.939\\
${M_{bol2}}(mag)$ &  $5.246\pm0.072$ & 5.173  \\ 
${L_1}({L_ \odot })$ & $1.832\pm0.121$ & 2.580\\ 
${L_2}({L_ \odot })$ &  $0.628\pm0.041$ & 0.668  \\ 
${R_1}({R_ \odot })$ & $1.430\pm0.049$ & 1.427\\
${R_2}({R_ \odot })$ & $0.749\pm0.026$ &   0.757\\
$a({R_ \odot })$ & $2.674\pm0.080$ & 2.64\\
$M_1({M_\odot})$ & $1.448\pm0.131$ & $1.467$\\
$M_2({M_\odot})$ & $0.362\pm0.037$ & $0.357$\\
$log(g)_1(cgs)$ & $4.288\pm0.008$ & 4.295\\
$log(g)_2(cgs)$ & $4.248\pm0.012$ & 4.232\\
\hline
\hline
\label{tab2}
\end{tabular}
\end{center}
\end{table}
\vspace{1.5cm}
\section{The Orbital Period Changes}
To calculate the eclipse times of minima, we used the same method as done by \citet{2020NewA...8001394S}. First, split the detrended light curves for individual eclipses and fit a Lorentzian function to each eclipse by using the least-squares method. We used the \texttt{Scipy.curve-fit} package in Python to fit the Lorentzian function to the individual eclipses. We used the \texttt{np.sqrt(np.diag(cov))} function to calculate the standard deviation errors on the parameters. TESS observations yielded a total of 220 primary and secondary minima, as displayed in the table \ref{tab3}. The new observational eclipse times were calculated. Then, we performed an analysis of observed (O) minus calculated (C) eclipse times \citep{2005ASPC..335....3S}. We calculated the O-C curve using the following linear ephemeris (\citealt{2004AcA....54..207K}, \citealt{2015PASJ...67...42Y}):

\begin{equation}
  \label{eq:seven}
Min.I = 2452500.1757 + 0.3765545 \times E
\end{equation}

\begin{figure*}
\begin{center}
  \includegraphics[scale=0.40]{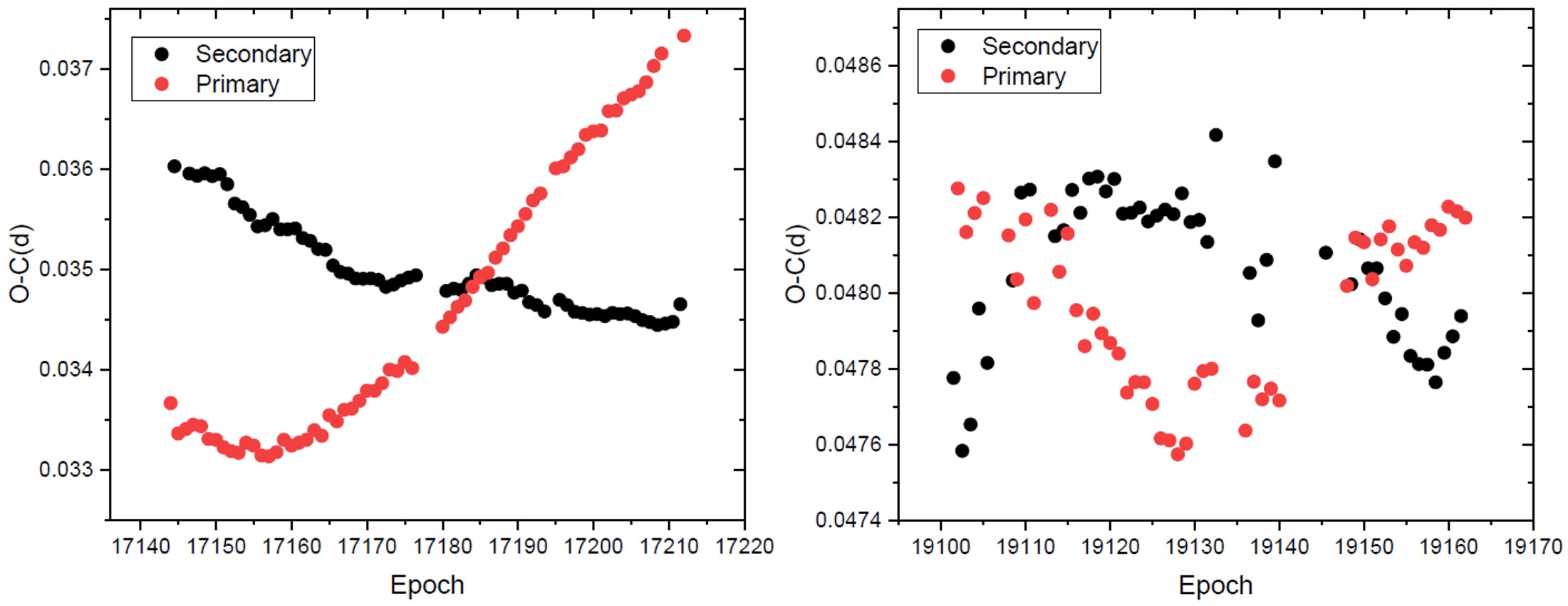}
  \caption{Left panel: primary and secondary O-C curve of minima for sector 24. Right panel: primary and secondary minima for sector 51 (primary O-C curve in black circles and secondary O-C curve in red circles.).}
  \label{Fig5}
  \end{center}
\end{figure*}

\begin{figure*}
\begin{center}
  \includegraphics[scale=0.90]{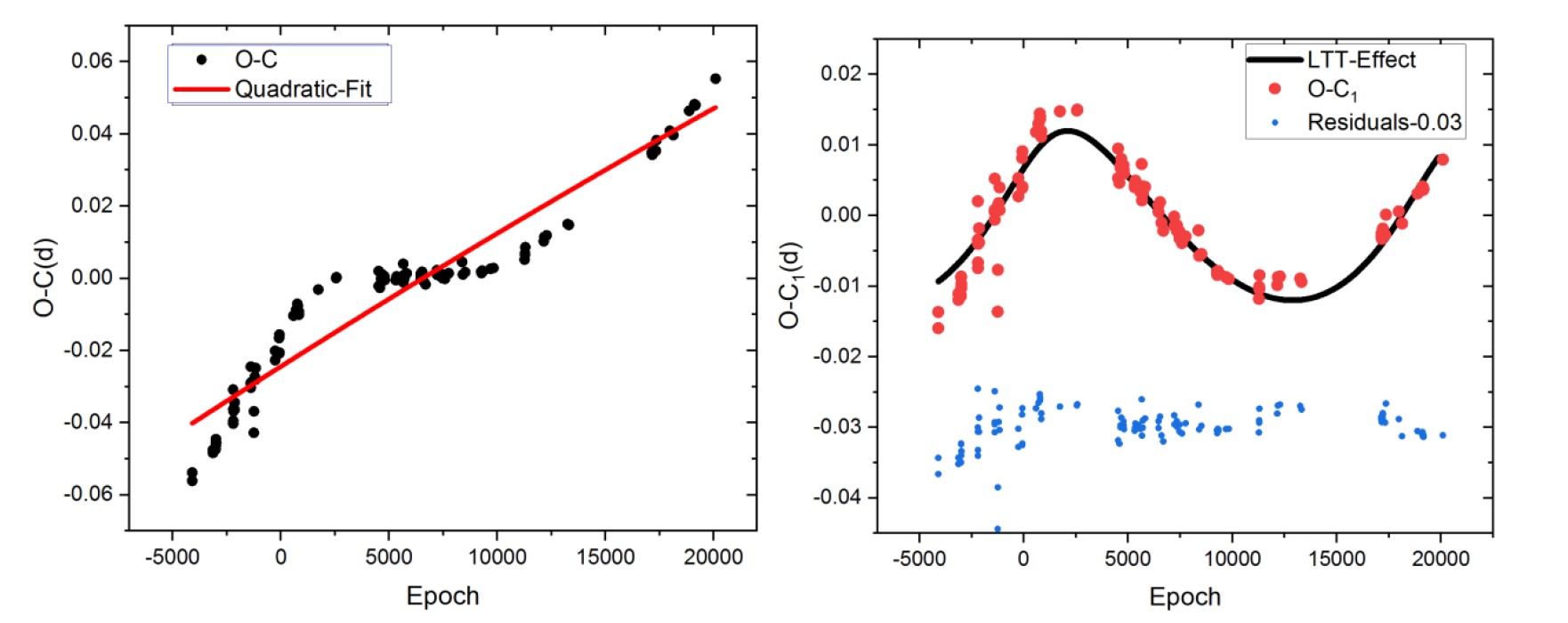}
  \caption{Left Panel: The O-C data points of the minima (black circle) and, the polynomial fit (red line). Right Panel: LTTE on the residuals of the polynomial fit; LTTE (black line), the O-C data points after subtracted polynomial fit (red circles), and the residuals of the LTT effect fit (blue circles).}
  \label{Fig6}
  \end{center}
\end{figure*}

The O-C curve of primary and secondary minima for sectors 24 and 51 are plotted in Figure \ref{Fig5}. The anti-correlated manner between primary and secondary minima is obvious which is a confirmation of the presence of spots on the contact binary components (\citet{2013ApJ...774...81T}; \citet{2015MNRAS.448..429B}). We averaged primary and secondary minima to eliminate the anti-correlated impact when analyzing orbital period changes  \citep{2015MNRAS.448..429B}. The calculated values are shown in Table \ref{tab3}. 109 YY CrB observational minima times were recorded in the literature over a 31-year period. The appendix contains a list of the data gathered with uncertainty. Observational minima times converted to BJD-TDB\footnote{\url{https://astroutils.astronomy.osu.edu/time/hjd2bjd.html}}.
Figure \ref{Fig6}'s left panel depicts the O-C curve of the minima. We presented a new ephemeris for this target as Equation ~(\ref{eq:eight}) by fitting a linear function on the O-C curve of primary.

\begin{equation}
  \label{eq:eight}
Min.I = 2458955.8598( \pm 1.1e - 4) + 0.3765581( \pm 1.1e - 7) \times E
\end{equation}
The O-C curve of minima is calculated with the new ephemeris and the resulting curve shows the same shape as the left panel of Figure \ref{Fig6}. 
we fitted a quadratic function to the O-C curve in order to investigate the variations in the orbital period:

\begin{equation}
    {T_{mid}}(E) = {T_0} + PE + \frac{1}{2}\frac{{dP}}{{dt}}{E^2}
\end{equation}

where mid-eclipse times ${T_{mid}}$ are described by ${T_0}$ is the reference mid-eclipse time, P is the orbital period and, E is the epoch of eclipses \citep{2017AJ....154....4P}. And the quadratic fit showed a drop in period, which corresponded to the quadratic plot in Figure \ref{Fig6}'s left panel. We determined the rate of period decline using the model's quadratic coefficient as Equation ~(\ref{eq:ten})

\begin{equation}
  \label{eq:ten}
\mathop P\limits^.  = \frac{{2 \times   -1.124 \times {{10}^{ - 11}}}}{{0.3765545}} =   -5.786 \times {10^{ - 8}} \pm 9.965 \times {10^{ - 9}}\frac{{day}}{{year}}
\end{equation}

 Considering ${M_1} = 1.448{M_ \odot }$ for the primary and ${M_2} = 0.362{M_ \odot }$ for the secondary one calculated in this study and using Equation ~(\ref{eq:elevn}) and mass conservation, the rate of mass exchange between primary and secondary components was estimated.

\begin{equation}
\label{eq:elevn}
    \frac{{\mathop P\limits^. }}{P} =  - 3\frac{{{{\mathop M\limits^. }_2}({M_1} - {M_2})}}{{{M_1}{M_2}}} \Rightarrow \mathop {{M_2}}\limits^.  = +2.472 \times {10^{ - 8}} \pm 0.190 \times {10^{ - 8}}{M_ \odot }y{r^{ - 1}}                                                            
\end{equation}
The positive sign indicates the direction of mass transfer from the more massive to the less massive component. The cyclic changes are shown by the residuals of the quadratic fit. As a result, we investigated the Light Travel Time Effect (LTTE) as a possible cause of the O-C curve variations. The following periodogram analysis was performed with the Period 04 software \citep{2005CoAst.146...53L} for the residuals of a quadratic fit. The peak of frequencies in the periodogram analysis of residuals shows a period of 7351.018 days. Then, we used the least square method to fit the Light Travel Time (LTT) formula on the O-C curve \citep{1952ApJ...116..211I}:
\begin{equation}
\label{eq:twelve}
    {(O - C)_{LTT}} = A \times \left( {\frac{{1 - {e^2}}}{{1 + e\cos \upsilon }}\sin (\upsilon  + \omega ) + e\sin (\omega )} \right)                                                                                                                                                 
\end{equation}

where $A = \frac{{{a_{12}}\sin i}}{c}$, and ${a_{12}}$ is the semi-major axis of the relative orbit of the eclipsing system around the center of mass (in Au unit), $i$ is the inclination of the third-body orbit, $e$ is the eccentricity of the supposed third body, $\omega$ is the longitude of periastron passage in the plane of the orbit and, $\upsilon$  is the true anomaly. To fit the LTT function to the residuals of the O-C curve, we have to convert the epoch to the true anomaly and Kepler's formula provides the link between the eccentric anomaly and the observed eclipse time: 
\begin{equation}
\label{eq:threen}
    {E_3} - e\sin E{}_3 = \frac{{2\pi }}{{{P_3}}}(t - {T_0})                                   
\end{equation}

The equation ~(\ref{eq:threen}) was calculated using Newton-Raphson's method for every eclipse time of minima and considering equation ~(\ref{eq:fourteen}) the epochs converted to the true anomaly.

\begin{equation}
\label{eq:fourteen}
\begin{array}{l}\tan \frac{\nu }{2} = {(\frac{{1 + e}}{{1 - e}})^{1/2}}\tan \frac{{{E_3}}}{2}\\t = {T_0} + epoch \times {P_{binary}}\end{array}                                      \end{equation}   

where ${P_binary}$, t, ${P_3}$, ${E_3}$, and ${T_0}$ are the period of binary, the time of observed minima, the period of the third body, the eccentric anomaly and, the time of periastron passage respectively. By assuming a coplanar orbit ($i=90$), we determined the lower limit for the mass of the third body. The calculated parameters of the third body are listed in Table \ref{Tab4} and the related curve is plotted in the right panel of Figure \ref{Fig6}.

\begin{table*}
\caption{Extracted the times of minima from TESS observations. All times of minima have been reduced to 2450000. Additionally, the minimum error is equal to 0.0001.}
\centering
\begin{center}
\footnotesize
\begin{tabular}{c c c c c c c c c c c c}
 \hline
 \hline
Min. & Epoch & O-C & Min. & Epoch & O-C & Min. & Epoch & O-C & Min. & Epoch & O-C\\
\hline
8955.8597	&	17144	&	0.0337	&	8966.5927	&	17172.5	&	0.0348	&	8978.8326	&	17205	&	0.0368	&	9702.0170	&	19125.5	&	0.0482	\\
8956.0504	&	17144.5	&	0.0360	&	8966.7802	&	17173	&	0.0340	&	8979.0187	&	17205.5	&	0.0345	&	9702.2047	&	19126	&	0.0351	\\
8956.2360	&	17145	&	0.0336	&	8966.9693	&	17173.5	&	0.0348	&	8979.2092	&	17206	&	0.0368	&	9702.2047	&	19126	&	0.0476	\\
8956.6126	&	17146	&	0.0334	&	8967.1567	&	17174	&	0.0340	&	8979.3952	&	17206.5	&	0.0344	&	9702.3936	&	19126.5	&	0.0482	\\
8956.8034	&	17146.5	&	0.0359	&	8967.3459	&	17174.5	&	0.0349	&	8979.5859	&	17207	&	0.0369	&	9702.5813	&	19127	&	0.0476	\\
8956.9892	&	17147	&	0.0334	&	8967.5334	&	17175	&	0.0341	&	8979.7718	&	17207.5	&	0.0345	&	9702.7702	&	19127.5	&	0.0482	\\
8957.1799	&	17147.5	&	0.0359	&	8967.7225	&	17175.5	&	0.0349	&	8979.9626	&	17208	&	0.0370	&	9702.9578	&	19128	&	0.0475	\\
8957.3657	&	17151	&	0.0002	&	8967.9099	&	17176	&	0.0340	&	8980.1483	&	17208.5	&	0.0344	&	9703.1468	&	19128.5	&	0.0482	\\
8957.3657	&	17148	&	0.0334	&	8968.0990	&	17176.5	&	0.0349	&	8980.3392	&	17209	&	0.0371	&	9703.3343	&	19129	&	0.0476	\\
8957.5565	&	17148.5	&	0.0359	&	8969.4165	&	17180	&	0.0344	&	8980.5249	&	17209.5	&	0.0344	&	9703.5233	&	19129.5	&	0.0482	\\
8957.7421	&	17149	&	0.0333	&	8969.6050	&	17180.5	&	0.0348	&	8980.9014	&	17210.5	&	0.0345	&	9703.7111	&	19130	&	0.0477	\\
8957.9331	&	17149.5	&	0.0359	&	8969.7931	&	17181	&	0.0345	&	8981.2781	&	17211.5	&	0.0346	&	9703.8998	&	19130.5	&	0.0482	\\
8958.1187	&	17150	&	0.0333	&	8969.9817	&	17181.5	&	0.0348	&	8981.4691	&	17212	&	0.0373	&	9704.0876	&	19131	&	0.0478	\\
8958.3096	&	17150.5	&	0.0359	&	8970.1697	&	17182	&	0.0346	&	9692.9793	&	19101.5	&	0.0478	&	9704.2763	&	19131.5	&	0.0481	\\
8958.4952	&	17151	&	0.0332	&	8970.3582	&	17182.5	&	0.0348	&	9693.1681	&	19102	&	0.0482	&	9704.4642	&	19132	&	0.0478	\\
8958.6861	&	17151.5	&	0.0358	&	8970.5464	&	17183	&	0.0347	&	9693.3556	&	19102.5	&	0.0475	&	9704.6531	&	19132.5	&	0.0484	\\
8958.8717	&	17152	&	0.0332	&	8970.7348	&	17183.5	&	0.0348	&	9693.5445	&	19103	&	0.0481	&	9705.9703	&	19136	&	0.0476	\\
8959.0624	&	17152.5	&	0.0356	&	8970.9231	&	17184	&	0.0348	&	9693.7323	&	19103.5	&	0.0476	&	9706.1590	&	19136.5	&	0.048	\\
8959.2482	&	17153	&	0.0332	&	8971.1114	&	17184.5	&	0.0349	&	9693.9211	&	19104	&	0.0482	&	9706.3469	&	19137	&	0.0478	\\
8959.4390	&	17153.5	&	0.0356	&	8971.2997	&	17185	&	0.0349	&	9694.1092	&	19104.5	&	0.0479	&	9706.5354	&	19137.5	&	0.0479	\\
8959.6249	&	17154	&	0.0333	&	8971.4880	&	17185.5	&	0.0349	&	9694.2977	&	19105	&	0.0482	&	9706.7234	&	19138	&	0.0477	\\
8959.8155	&	17154.5	&	0.0355	&	8971.6763	&	17186	&	0.0349	&	9694.4855	&	19105.5	&	0.0478	&	9706.9121	&	19138.5	&	0.0481	\\
8960.0014	&	17155	&	0.0332	&	8971.8645	&	17186.5	&	0.0348	&	9695.2398	&	19107.5	&	0.0489	&	9707.1000	&	19139	&	0.0477	\\
8960.1919	&	17155.5	&	0.0354	&	8972.0531	&	17187	&	0.0351	&	9695.4272	&	19108	&	0.0481	&	9707.2889	&	19139.5	&	0.0483	\\
8960.3778	&	17156	&	0.0331	&	8972.2411	&	17187.5	&	0.0348	&	9695.6154	&	19108.5	&	0.048	&	9707.4766	&	19140	&	0.0477	\\
8960.5685	&	17156.5	&	0.0354	&	8972.4297	&	17188	&	0.0352	&	9695.8037	&	19109	&	0.0480	&	9708.2306	&	19142	&	0.0486	\\
8960.7544	&	17157	&	0.0331	&	8972.6176	&	17188.5	&	0.0348	&	9695.9922	&	19109.5	&	0.0483	&	9709.5480	&	19145.5	&	0.0481	\\
8960.9450	&	17157.5	&	0.0355	&	8972.8063	&	17189	&	0.0353	&	9696.1804	&	19110	&	0.0482	&	9710.4893	&	19148	&	0.048	\\
8961.1310	&	17158	&	0.0331	&	8972.9941	&	17189.5	&	0.0348	&	9696.3688	&	19110.5	&	0.0483	&	9710.6776	&	19148.5	&	0.0480	\\
8961.3215	&	17158.5	&	0.0354	&	8973.1830	&	17190	&	0.0354	&	9696.5567	&	19111	&	0.0479	&	9710.8660	&	19149	&	0.0481	\\
8961.5077	&	17159	&	0.0333	&	8973.3707	&	17190.5	&	0.0348	&	9697.3101	&	19113	&	0.0482	&	9711.0543	&	19149.5	&	0.0481	\\
8961.6980	&	17159.5	&	0.0354	&	8973.5597	&	17191	&	0.0355	&	9697.4983	&	19113.5	&	0.0481	&	9711.2425	&	19150	&	0.0481	\\
8961.8842	&	17160	&	0.0332	&	8973.7471	&	17191.5	&	0.0347	&	9697.6865	&	19114	&	0.0480	&	9711.4307	&	19150.5	&	0.0481	\\
8962.0746	&	17160.5	&	0.0354	&	8973.9364	&	17192	&	0.0357	&	9697.8749	&	19114.5	&	0.0482	&	9711.6190	&	19151	&	0.048	\\
8962.2608	&	17161	&	0.0331	&	8974.1236	&	17192.5	&	0.0346	&	9697.8749	&	19114.5	&	0.0482	&	9711.8073	&	19151.5	&	0.0481	\\
8962.4511	&	17161.5	&	0.0353	&	8974.3130	&	17193	&	0.0375	&	9698.0631	&	19115	&	0.0481	&	9711.9956	&	19152	&	0.0481	\\
8962.6374	&	17162	&	0.0333	&	8974.5001	&	17193.5	&	0.0345	&	9698.2516	&	19115.5	&	0.0483	&	9712.1838	&	19152.5	&	0.048	\\
8962.8276	&	17162.5	&	0.0353	&	8975.0664	&	17195	&	0.0360	&	9698.4395	&	19116	&	0.0479	&	9712.3722	&	19153	&	0.0482	\\
8963.0140	&	17163	&	0.0334	&	8975.2533	&	17195.5	&	0.0347	&	9698.6281	&	19116.5	&	0.0482	&	9712.5602	&	19153.5	&	0.0479	\\
8963.2041	&	17163.5	&	0.0352	&	8975.4430	&	17196	&	0.0360	&	9698.8160	&	19117	&	0.0478	&	9712.7487	&	19154	&	0.0481	\\
8963.3905	&	17164	&	0.0333	&	8975.6298	&	17196.5	&	0.0346	&	9699.0047	&	19117.5	&	0.0483	&	9712.9368	&	19154.5	&	0.0479	\\
8963.5807	&	17164.5	&	0.0351	&	8975.8196	&	17197	&	0.0361	&	9699.1926	&	19118	&	0.0479	&	9713.1252	&	19155	&	0.0481	\\
8963.7673	&	17165	&	0.0335	&	8976.0063	&	17197.5	&	0.0346	&	9699.3812	&	19118.5	&	0.0483	&	9713.5019	&	19156	&	0.0481	\\
8963.9570	&	17165.5	&	0.035	&	8976.1962	&	17198	&	0.0362	&	9699.5691	&	19119	&	0.0479	&	9713.6898	&	19156.5	&	0.0478	\\
8964.1438	&	17166	&	0.0335	&	8976.3828	&	17198.5	&	0.0346	&	9699.7578	&	19119.5	&	0.0482	&	9713.8784	&	19157	&	0.0481	\\
8964.3335	&	17166.5	&	0.035	&	8976.5729	&	17199	&	0.0363	&	9699.9456	&	19120	&	0.0479	&	9714.0664	&	19157.5	&	0.0478	\\
8964.5204	&	17167	&	0.0336	&	8976.7594	&	17199.5	&	0.0345	&	9700.1344	&	19120.5	&	0.0483	&	9714.2550	&	19158	&	0.0482	\\
8964.7101	&	17167.5	&	0.0349	&	8976.9495	&	17200	&	0.0364	&	9700.3221	&	19121	&	0.0478	&	9714.4429	&	19158.5	&	0.0477	\\
8964.8970	&	17168	&	0.0336	&	8977.1360	&	17200.5	&	0.0345	&	9700.3222	&	19122	&	0.004	&	9714.6315	&	19159	&	0.0481	\\
8965.0865	&	17168.5	&	0.0349	&	8977.3260	&	17201	&	0.0009	&	9700.5108	&	19121.5	&	0.0482	&	9714.8195	&	19159.5	&	0.0478	\\
8965.2736	&	17169	&	0.0337	&	8977.5125	&	17201.5	&	0.0345	&	9700.6986	&	19122	&	0.0477	&	9715.0082	&	19160	&	0.0482	\\
8965.4631	&	17169.5	&	0.0349	&	8977.7028	&	17202	&	0.0364	&	9700.8874	&	19122.5	&	0.0482	&	9715.1961	&	19160.5	&	0.0478	\\
8965.6503	&	17170	&	0.0338	&	8977.8891	&	17202.5	&	0.0345	&	9701.0752	&	19123	&	0.0478	&	9715.1961	&	19160.5	&	0.0478	\\
8965.8397	&	17170.5	&	0.0349	&	8978.0794	&	17203	&	0.0366	&	9701.2639	&	19123.5	&	0.0482	&	9715.3847	&	19161	&	0.0482	\\
8966.0269	&	17171	&	0.0338	&	8978.2656	&	17203.5	&	0.0345	&	9701.4517	&	19124	&	0.0478	&	9715.5727	&	19161.5	&	0.0479	\\
8966.2162	&	17171.5	&	0.0349	&	8978.4560	&	17204	&	0.0367	&	9701.6405	&	19124.5	&	0.0481	&	9715.7612	&	19162	&	0.0482	\\
8966.4035	&	17172	&	0.0339	&	8978.6422	&	17204.5	&	0.0345	&	9701.8282	&	19125	&	0.0476	&		&		&		\\
\hline
\hline
\end{tabular}
\end{center}
\label{tab3}
\end{table*}
\begin{table}
\centering
\begin{center}
\caption{Parameters of the third body.}
\footnotesize
\begin{tabular}{c c c}
 \hline
 \hline
Parameters of third body && Value \\
\hline
Eccentricity (e) && $0.689 \pm 0.005$ \\
The longitude of periastron passage ($\omega$) &&  $57.6 \pm 1.8$   \\ 
Period (days)  && 7351.018\\ 
Amplitude (minutes) &&  $17.28 \pm 1.44$  \\ 
The time of periastron passage ($T_{0}$) && 2460201\\
Projected semi-major axis$ \times \sin i$ && $2.11 \pm 0.17$ \\
Mass Function $(Mass Function ({f_m}))$&& $0.023 \pm 0.006$\\
${M_3}\sin i (i = 90,{M_ \odot })$ && 0.498\\
$\sum {{{(O - C)}^2}} $&& 0.001\\
\hline
\hline
\end{tabular}
\end{center}
\label{Tab4}
\end{table}
Assuming the third body is a main-sequence star, this corresponds to the M1V spectral type with a brightness of $0.041{L_{\odot}}$ \footnote {\url{http://www.pas.rochester.edu/~emamajek/EEM_dwarf_UBVIJHK_colors_Teff.txt}}
, or $0.016$ of the total luminosity which doesn't agree with the value $l_3$ determined in section 3.

We are not certain that $l_3$ produced from the TESS light curve analysis represents a valid observation of flux from a third body in the system, despite the computation of the effect of the third body, especially as there is no exact radial-velocity curve. We estimate that this figure is most likely due to systematic errors in background flux level readings in TESS photos and/or an underestimation of photometric aperture contamination by other stars in the image. So, we investigated the Applegate's effect as a plausible explanation for the cyclical fluctuations in the O-C curve.  We calculated the observed relative change of the orbital period throughout one cycle of the binary using the previously obtained modulation period and the O-C amplitude computed from the orbit of the third body simulated in the previous section. $\frac{{\Delta P}}{P} = 2\pi \frac{{(O - C)}}{{{P_{\bmod }}}} = 1.025 \times {10^{ - 5}}$ \citep{1992ApJ...385..621A}. 
The value of $\frac{{\Delta P}}{P}$suggests that the Applegate effect can explain the cyclic changes in the O-C curve of minima.

This system shows unequal maxima that is known as the O'Connell effect \citep{1951PRCO....2...85O} due to the presence of the hot spots \citep{2009SASS...28..107W}. So, this difference implies that the hemisphere of a component emits a different amount of radiation than the other hemisphere. These types of systems have active chromospheres because of the existence of large spots \citep{2022ApJS..262...10K}. Starspots can alter the depth of minima and have an obvious effect on the eclipse light curve \citep{2019AJ....158..111H}.

\begin{figure*}
  \begin{center}

  \includegraphics[scale=0.40]{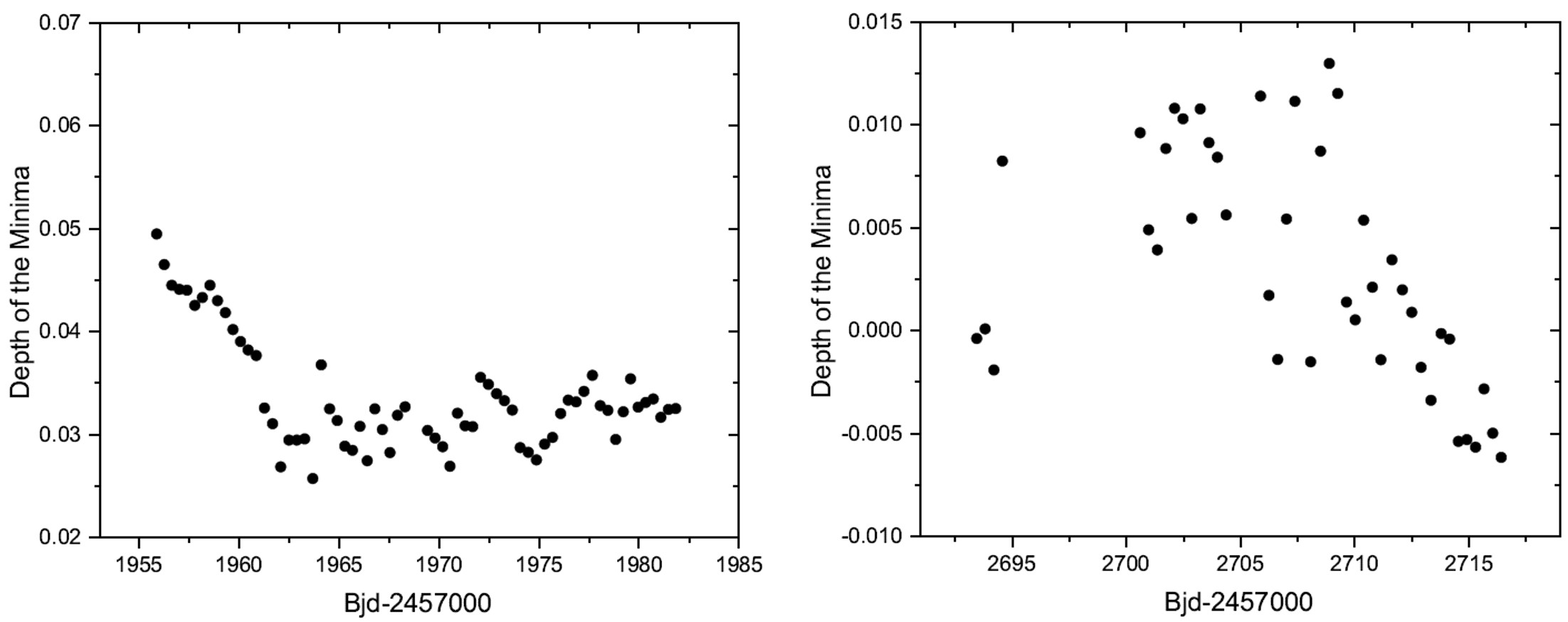}
  \caption{Right panel: the difference between the depths of Primary and Secondary minima over the 24th sector of TESS observations. Left panel: the difference between the depths of Primary and Secondary minima over the 51st sector of TESS observations.}
  \label{Fig7}
 \end{center}
\end{figure*}

To investigate the effect of the spots on the light curve over both sectors of TESS observations, we calculated the difference between the two depths of the primary and secondary minima. We considered the relative fluxes in phases 0 and 0.5 for every complete individual light curve. The curves that resulted were displayed in Figure \ref{Fig7}. YY CrB has the values of the DepthI - DepthII as large as about $10\% $  of the variable light amplitude. And this is possible because of the migration and evolution of spots with time on the surface of two components that cause cyclic magnetic activity.
\vspace{1cm}
\section{Discussion and conclusion}
\begin{figure}
\begin{center}
  \includegraphics[scale=0.60]{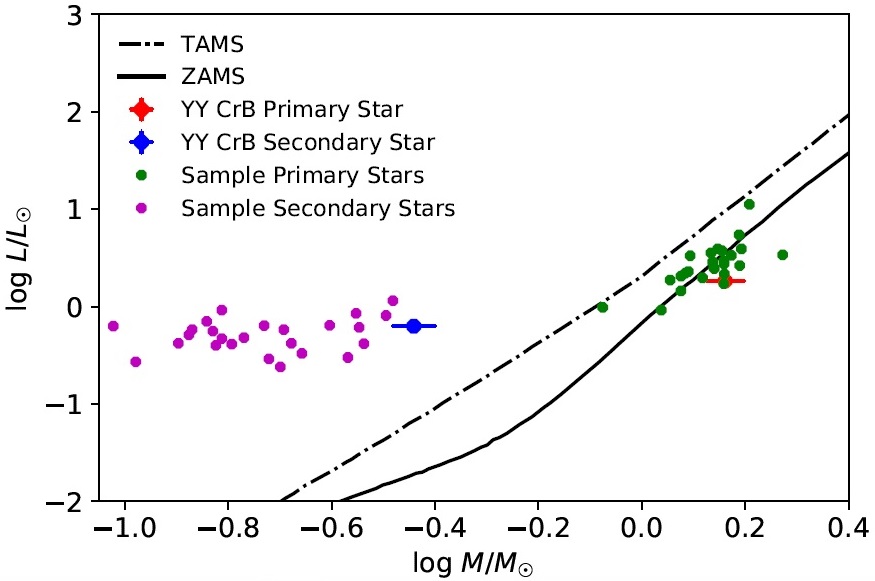}
  \caption{$M-L$ diagram of selected contact binaries with low mass ratio. The primary and secondary components of YY CrB are plotted in red and blue colors, respectively.}
  \label{Fig8}
  \end{center}
\end{figure}

\begin{figure}
\begin{center}
\includegraphics[scale=0.60]{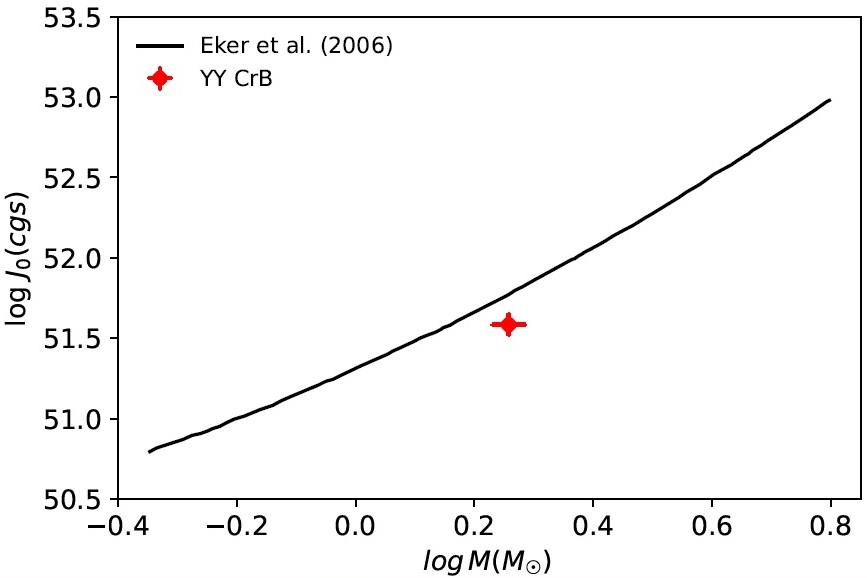}
    \caption{The location of YY CrB on the $logJ_0-logM$ diagram. The quadratic line is based on a study by \citet{2006MNRAS.373.1483E}.}
\label{Fig9}
\end{center}
\end{figure}

Based on the estimated mass ratio, fill-out factor and inclination angle, YY CrB is an over-cantact binary with an increased orbital period. \citet{2010NewA...15..227E} calculated the decreased period rate $1.194 \times {10^{ - 6}}\frac{{day}}{{year}}$. \citet{2015PASJ...67...42Y} considered all of the minima time published until 2013 and calculated a secular period decrease with a rate of $6.727 \times {10^{ - 7}}\frac{{day}}{{year}}$. In this study, the decreasing value of period rate $5.786 \times {10^{ - 8}}\frac{{day}}{{year}}$ indicates that the rate of period changes has been decreased.
And, when mass conservation is taken into account, the mass transfer from the Roche-lobe-filling primary component to the secondary component is ${\mathop M\limits^. _2} = 2.472 \times {10^{ - 8}}{M_ \odot }y{r^{ - 1}}$. When the stated results in \citet{2015PASJ...67...42Y} are compared to the value of mass transfer in this study, it is clear that mass transfer has been decreased and the distance between two components is growing while the value of fill-out factor is decreasing (compare the values of fill-out factors in Table \ref{Tab1}). And this target may evolve to shallow-contact binary via the thermal relaxation oscillation (TRO) model (\citet{1976ApJ...205..217F}; \citet{1977MNRAS.179..359R}) and ultimately reach a broken-contact phase ( \citet{1976ApJ...205..208L}).
The mass ratio of the components, which is related to the mass transfer, is the crucial parameter in the evolution of the close binary stars. Table \ref{Tab5} contains a list of contact binaries with low mass ratios $<0.25$. In order to explain the evolutionary status of the YY CrB system, we provide the mass-luminosity ($M-L$) diagram displayed in Figure \ref{Fig8}. The Zero-Age Main Sequence (ZAMS) and the Terminal‐Age Main Sequence (TAMS) are plotted along with the selected contact binaries with low mass ratios. It is obvious that the more massive primary components are around the ZAMS line, meaning they are not evolved or little evolved. Also, the less massive secondary components have evolved away from the main sequence stars and over-luminosity comparing the stars with the same mass in the main sequence. In addition, the orbital angular momentum of YY CrB has a value of $51.585\pm0.067$. The $logJ_0-logM$ diagram shows the position of the system (Figure \ref{Fig9}), and this diagram shows that YY CrB is in a contact binary systems region.
\begin{table*}
\caption{Absolute parameters for low mass ratio contact binaries.}
\centering
\begin{center}
\footnotesize
\begin{tabular}{c c c c c c c c c}
 \hline
 \hline
System & $q$ & ${M_1}({M_\odot})$ & ${M_2}({M_\odot})$ & ${R_1}({R_\odot})$ & ${R_2}({R_\odot})$ & ${L_1}({L_\odot})$ & ${L_2}({L_\odot})$ & Reference \\
\hline
V429 Cam	&	0.206	&	1.36(12)	&	0.28(3)	&	1.55(3)	&	0.78(2)	&	3.56(9)	&	0.85(2)	&	\citet{2021AJ....162...13L} \\
V830 Cep 	&	0.23	&	0.84(5)	&	0.19(1)	&	0.91(1)	&	0.47(1)	&	0.98(1)	&	0.29(1)	&	\citet{2021AJ....162...13L} \\ 
FP Boo	&	0.096	&	1.614(52)	&	0.154(21)	&	2.310(25)	&	0.774(8)	&	11.193(99)&	0.920(13)	&	\citet{2006AcA....56..127G}\\ 
DN Boo	&	0.103	&	1.428(39)	&	0.148(6)	&	1.710(67)	&	0.670(110)	&	3.750(280)	&	0.560(170)	&	 \citet{2008NewA...13..468S} \\ 
FG Hya	&	0.112	&	1.444(25)	&	0.161(7)	&	1.405(9)	&	0.591(8)	&	2.158(86)	&	0.412(17)	&	\citet{2005MNRAS.356..765Q}\\
CK Boo	&	0.0108	&	1.442(14)	&	0.154(2)	&	1.453(3)	&		0.577(10)&	2.74(1)&	0.47(2)	&	\citet{2005AN....326..342K} \\
GR Vir	&	0.122	&	1.37(16)	&	0.17(6)	&	1.42(7)	&	0.61(4)	&	2.87(28)	&	0.48(6)	&	\citet{2004AJ....128.2430Q}\\
CSS J234807.2$+$193717	&	0.176	&	1.19(4)&	0.21(3)	&	1.36 (2)	&	0.66 (1)	&	1.45(24)&	0.42(5)	&	\citet{2022MNRAS.512.1244C}\\
J170307&	0.092&	1.134(253)&	0.105(24)	&	1.204(120)&	0.436(48)&	1.874(572)&	0.271(72)&	\citet{2023MNRAS.519.5760L}\\
J1641000&	0.095&	1.402(287)&	0.133(28)&	1.580(144)&	0.577(58)&	3.912(1.109)&	0.512(142)&	\citet{2023MNRAS.519.5760L}\\
J223837&	0.093&	1.541(306)&	0.144(30)&	1.784(159)&	0.646(64)&	5.463(1.534)&	0.704(192)&	\citet{2023MNRAS.519.5760L}\\
CSS J222607.8$+$062107&	0.221&	1.49(3)&	0.33(13)&	1.51(5)&	0.81(2)&	3.35(65)&	1.15(24)& \citet{2020ApJS..247...50S}\\
CSS J012559.7$+$203404&	0.231&	1.38(3)&	0.32(12)&	1.42(4)&	0.77(2)&	2.46(56)&	0.81(18)&	\citet{2020ApJS..247...50S}\\\
CSS J153855.6$+$042903&	0.187&	1.44(5)&	0.27(12)&	1.37(4)&	0.66(2)&	2.94(96)&	0.30(10)&	\citet{2020ApJS..247...50S}\\\
CSS J141923.2$-$013522&	0.168&	1.31(5)&	0.22(11)&	1.23(4)&	0.57(2)&	1.97(68)&	0.33(11)&	\citet{2020ApJS..247...50S}\\\
CSS J130111.2$-$132012&	0.108&	1.38(3)&	0.15(12)&	1.49(5)&	0.61(2)&	2.49(57)&	0.40(9)&	\citet{2020ApJS..247...50S}\\\
CSS J165813.7$+$390911 &	0.183&	1.09(3)&	0.20(9)&	1.05(3)&	0.49(1)&	0.92(24)&	0.24(6)&	\citet{2020ApJS..247...50S}\\\
V870 Ara&	0.082&	1.546(54)&	0.127(37)&	1.64(6)&	0.63(5)&	2.64(17)&	0.42(6)&	\citet{2021OAst...30...37P}\\
TYC 6995-813-1&	0.111&	1.23(1)&	0.135(1)&	1.46(1)&	0.60(1)&	2.293(4)&0.58(2)&	\citet{2021RAA....21..235W} \\
NSVS 13602901&	0.171&	1.19(2)&	0.203(10)&	1.69(1)&	0.79(1)&	2.05(4)&	0.58(2)&	\citet{2021RAA....21..235W} \\
NSVS 5029961&	0.151&	1.872(468)&	0.284(72)&	1.573(119)&	0.680(53)&	3.403(14)&	0.610(26)&	\citet{2021MNRAS.506.4251Z} \\
CSS J022914.4$+$044340&	0.201&	1.44(25)&	0.29(5)&	1.26(8)&	0.65(4)&	1.718(191)&	0.416(50)&	\citet{2021RAA....21..180L}\\
HV Aqr&	0.15&	1.240(28)&	0.186(17)&	1.456(12)&	0.601(5)&	3.326(213)&	0.638(44)&	\citet{2021MNRAS.501.2897G} \\
ZZ PsA&	0.078&	1.213(8)&	0.095(1)&	1.422(4)&	0.559(4)&	2.20(4)&	0.63(4)&	\citet{2021MNRAS.501..229W}\\
NSVS 1926064&	0.160&	1.558(38)&	0.249(6)&	1.605(13)&	0.755(42)&	3.91(28)&	0.641(33)&	\citet{2020NewA...7701352K} \\

\hline
\hline
\end{tabular}
\end{center}
\label{Tab5}
\end{table*}

According to the study \citet{2015PASJ...67...42Y} and the periodic changes in the residuals of the quadratic fit on the O-C curve, the potential of excitability of the third body was investigated. The light-time function fitting revealed the existence of a body with the value of $0.498{M_ \odot }$.
This number equates to 0.016 of total brightness, which differs from the value of $l_3$ determined in section 3. 
 We explored the Applegate effect as a possible explanation for the fluctuation in the O-C curve because there is no exact radial-velocity curve.

YY CrB is a contact binary with a mass ratio less than $0.3$, so considering Hut's criteria \citep{1980A&A....92..167H} to investigate the stability is necessary. We used Equation ~(\ref{eq:fifteenth}) to calculate the ratio of the spin angular momentum to the orbital angular momentum \citep{2015AJ....150...69Y}.

\begin{equation}
\label{eq:fifteenth}
\frac{{{J_s}}}{{{J_o}}} = \frac{{q+1 }}{q}[{({k_1}{r_1})^2} + {({k_2}{r_2})^2}q]
\end{equation}

where $r_{1}$ and $r_{2}$ are the relative radii for the primary and secondary components and ${k^2}_{1,2} = 0.06$ \citep{2006MNRAS.369.2001L}, are the dimensionless gyration radii. The calculated value of $\frac{{{J_s}}}{{{J_o}}} = 0.087$, which is less than the threshold value therefore this system is stable. 
This target shows a period increase which is attributed to mass transfer. According to the existing data and the analyses done in this study, the existence of a third body is unlikely for this system, and detailed spectroscopic and photometric observations over a longer length of time are required for the definitive determination.

\vspace{1.5cm}
\section*{Acknowledgements}
This manuscript has made use of data from the TESS mission. Funding for the TESS mission is provided by the NASA Science Mission Directorate. This research has made use of the SIMBAD and VIZIER databases, operated at CDS, Strasbourg, France. The time of minima data from the Variable Star Observers League in Japan (VSOLJ) websites proved invaluable to the assessment of potential period changes experienced by this variable star. The authors would like to thank Marco Brentel for his help. We are grateful to Ehsan Paki from the BSN project (\url{https://bsnp.info/}) for providing Figure 4 of this manuscript, which also shows the color-temperature scale.

\vspace{1.5cm}
\section*{ORCID iDs}
\noindent Somayeh Soomandar: 0000-0002-9520-9573\\
Atila Poro: 0000-0002-0196-9732\\

\vspace{1.5cm}
\appendix
\section{Available Minima Times}
The appendix table displays the minima times along with their error in the first column, the epochs in the third column, the O-C values in the fourth column, and the references in the final column.

\setcounter{table}{0}
\begin{table*}
\caption{Available mid-eclipse times of YY CrB system. ons. All minimum have been reduced to 2450000}
\centering
\begin{center}
\footnotesize
\begin{tabular}{c c c c c c c c}
 \hline
 \hline
Min.($BJD_{TDB}$) & Epoch & O-C & Reference & Min.($BJD_{TDB}$) & Epoch & O-C & Reference\\
\hline
955.8695(12)	&	-4101	&	-0.0562	&	 \citet{2001IBVS.5056....1P} 	&	4308.3899(2)	&	4802	&	-0.0005	&	\citet{2009IBVS.5898....1P}	\\
955.8718(6)	&	-4101	&	-0.0539	&	 \citet{2000AJ....120.1133R}	&	4564.2595	&	5481.5	&	0.0003	&	 \citet{Nagai+2009}\\
1318.4993&	-3138	&	-0.04847	&	 \citet{2010NewA...15..227E} 	&	4604.36215(1)	&	5588	&	-0.0001	&	 \citet{2009IBVS.5887....1Y}	\\
1318.5001	&	-3138	&	-0.0475	&	 \citet{2010NewA...15..227E}	&	4605.4917(2)	&	5591	&	-0.0002	&	 \citet{2009IBVS.5887....1Y}	\\
1361.4275(3)	&	-3024	&	-0.0475	&	 \citet{2000IBVS.4855....1K} 	&	4628.4657(2)	&	5652	&	0.0039	&	 \citet{2009IBVS.5898....1P}	\\
1361.4275(2)	&	-3024	&	-0.0474	&	 \citet{2000IBVS.4855....1K}	&	4632.4144(3)	&	5662.5	&	-0.0012	&	\citet{2009IBVS.5898....1P}\\
1368.3965(4)	&	-3005.5	&	-0.0447	&	 \citet{2000IBVS.4855....1K} 	&	4648.4191(30)	&	5706	&	0.0000	&	 \citet{2009IBVS.5898....1P}	\\
1368.3966(4)	&	-3005.5	&	-0.0446	&	 \citet{2000IBVS.4855....1K}	&	4648.4201(10)	&	5705	&	0.0009	&	\citet{2009IBVS.5874....1H}	\\
1370.4659(6)	&	-3000	&	-0.0463	&	 \citet{2000IBVS.4855....1K} 	&	4688.3352(3)	&	5811	&	0.0013	&	 \citet{2009IBVS.5887....1Y}	\\
1372.3494(3)	&	-2995	&	-0.0457	&	 \citet{2000IBVS.4855....1K}	&4931.4009(1)	&	6456.5	&	0.0011	&	 \citet{2011yCatp018591800H}		\\
1668.3359&	-2209	&	-0.0309&	 \citet{2010NewA...15..227E} 	&	4931.5883(1)	&	6457	&	0.0001	&	 \citet{2011yCatp018591800H}	\\
1669.4602	&	-2206	&	-0.0363	&	 \citet{2010NewA...15..227E}	&	4958.5136(20)	&	6528.5	&	0.0018	&	 \citet{2011yCatp018591800H}	\\
1670.3976&	-2203.5	&	-0.0403&	 \citet{2010NewA...15..227E}	&	4983.7401(3)	&	6595.5	&	-0.0009	&	 \citet{2009IBVS.5894....1D}	\\
1670.3984 &	-2203.5	&	-0.0395	&	 \citet{2010NewA...15..227E} 	&	5017.44086(3)	&	6685	&	-0.0017	&	 \citet{2009IBVS.5898....1P}	\\
1674.3548&	-2193	&	-0.0369	&	 \citet{2003IBVS.5380....1K} 	&	5213.6297(3)	&	7206	&	0.0022	&	 \citet{2011IBVS.5980....1P}	\\
1692.4299	&	-2144	&	-0.0365	&	 \citet{2010NewA...15..227E}	&	5219.6532(1)	&	7222	&	0.0009	&	 \citet{2011IBVS.5980....1P}\\
1692.4319 &	-2145	&	-0.0345	&	 \citet{2010NewA...15..227E} 	&	5261.8279(1)	&	7334	&	0.0014	&	 \citet{2011IBVS.5974....1D}	\\
1975.6050	&	-1392	&	-0.0304	&	 \citet{2003CoSka..33...38P}	&	5264.4630(3)	&	7341	&	0.0006	&	 \citet{2011IBVS.5980....1P}	\\
1975.6061(1)	&	-1392	&	-0.0293	&	 \citet{2001IBVS.5056....1P}	&	5311.3444(2)	&	7465.5	&	0.001	&	 \citet{2011IBVS.5980....1P}	\\
1975.6064(1)	&	-1393	&	-0.0290&	 \citet{2001IBVS.5056....1P}	&5311.5317(2)	&	7466	&	0.000	&	 \citet{2011IBVS.5980....1P}		\\
1975.6108(7)	&	-1392	&	-0.0246	&	 \citet{2002CoSka..32...79P}	&	5351.4463(1)	&	7572	&	-0.0002	&	 \citet{2012IBVS.6010....1H}	\\
2029.4398(7)	&	-1250	&	-0.0426	&	 \citet{2002CoSka..32...79P}	&5354.4587(2)	&	7580	&	-0.0002	&	 \citet{2011IBVS.5980....1P}\\
2031.5168(7)	&	-1244.5	&	-0.0369	&	 \citet{2002CoSka..32...79P})	&	5420.3573(3)	&	7755	&	0.0137	&	 \citet{2011IBVS.5980....1P})	\\
2045.4589	&	-1207.5	&	-0.0273	&	 \citet{2010NewA...15..227E}	&	5652.8828(30)	&	8372.5	&	0.0045	&	 \citet{2011IBVS.5960....1D}	\\
2060.3320	&	-1167	&	-0.0281	&	 \citet{2010NewA...15..227E}	&	5665.4931(3)	&	8406	&	0.0088	&	 \citet{2013IBVS.6044....1P}	\\
2060.3352 &	-1167 &	-0.0249	&	 \citet{2010NewA...15..227E} 	&	5705.4093(23)	&	8512	&	0.0016	&	 \citet{2012IBVS.6010....1H} \\
2400.1804(2)	&	-265.5	&	-0.0202	&	 \citet{2002IBVS.5341....1P}&	6011.5475(2)	&	9325	&	0.0051	&	 \citet{2013IBVS.6044....1P}	\\
2400.3660&	-264&	-0.0227&	\citet{2003IBVS.5380....1K} &55987.6371(2)	&	9261.5	&	0.0018	&	 \citet{2013IBVS.6044....1P}\\\
2469.4699(4)	&	-81.5	&	-0.017	&	\citet{2003IBVS.5364....1D}&	  5987.6371(2)	&	9261.5	&	0.0018	&	 \citet{2013IBVS.6044....1P}\\
2472.2898(2)	&	-74	&	-0.0209	& \citet{2015IBVS.6153....1P} &	5992.5319(2)	&	9274.5	&	0.0014	&	 \citet{2013IBVS.6044....1P}	\\
2473.4197(2)	&	-71	&	-0.0206	&	 \citet{2015IBVS.6153....1P}	&	6005.5237(2)	&	9309	&	0.0021	&	 \citet{2013IBVS.6044....1P}	\\
2473.4247(4)	&	-71	&	-0.0156	&	 \citet{2003IBVS.5364....1D} 	&6005.5237(2)	&	9309	&	0.0021	&	 \citet{2013IBVS.6044....1P}	\\
2500.1757&	0	&	0.000&	 \citet{2004AcA....54..207K} 	&	6149.3679(2)	&	9691	&	0.0026	&	 \citet{2013IBVS.6044....1P}\\
2719.3200&	582	&	-0.0104&	 \citet{Nagai+2004} 	&2456199.26169(2)	&	9823.5	&	0.0028	&	\citet{2013IBVS.6044....1P}	\\
2764.5082(23)	&	702	&	-0.0088	&	\citet{2005IBVS.5657....1H} &	6742.4439(14)	&	11266	&	0.00513	&	 \citet{2015IBVS.6149....1H}\\
2786.3500(2)	&	761	&	-0.0071	&	\citet{2003IBVS.5462....1A} &6749.4115(5)	&	11284.5	&	0.0066	&	\citet{2015IBVS.6149....1H} \\
2793.5038(2)	&	779	&	-0.0079	&	\citet{2003IBVS.5462....1A} 	&	6011.5483(2)	&	9325	&	0.0018	&	 \citet{2013IBVS.6044....1P}	\\
7074.9466(1)	&	12149	&	0.010	&	 \citet{2016IBVS.6164....1N}&	6754.4970(35)	&	11298	&	0.0086	&	 \citet{2015IBVS.6149....1H}	\\
2793.5042(21)	&	779	&	-0.0075	&	\citet{2005IBVS.5657....1H} &7074.9466(1)	&	12149	&	0.010	&	 \citet{2016IBVS.6164....1N}\\
2814.4011(1)	&	834.5	&	-0.0093	&	\citet{2003IBVS.5471....1S} &	7084.1734 &	12173.5	&	0.0114	&	 \citet{Nagai+2016}	\\
3151.0470	&	1728.5	&	-0.0032	&	\citet{Nagai+2005}&	7123.5238(28)	&	12278	&	-0.011	&	 \citet{2017IBVS.6196....1H})\\
3458.8835(2)	&	2546	&	0.000	&	 \citet{2006IBVS.5677....1D}	&6749.6002(8)	&	11285	&	0.0068	&	 \citet{2015IBVS.6149....1H}	\\
3466.41483(1)	&	2566	&	0.000	&	\citet{2005IBVS.5668....1P}  &	7513.0723	&	13312.5	&	0.0147	&	 \citet{Nagai+2017}	\\
4201.4509(16)	&	4518	&	0.0020	&	 \citet{2007IBVS.5802....1H}	&7489.5378(18)	&	13250	&	0.015	&	 \citet{2017IBVS.6196....1H}	\\
4201.6350(17)	&	4518.5	&	-0.0022	&	\citet{2007IBVS.5802....1H}&9024.3943 &	17326&	0.0353	&	 \citet{Nagai+2021}	\\
4245.5055(9)	&	4635	&	-0.0003	&	\citet{2007OEJV...74....1B} &9037.3884 &17360.5	&	0.0382	&	 \citet{Nagai+2021}	\\
4224.4161(2)	&	4579	&	-0.0028	&	 \citet{2009IBVS.5898....1P}	&9269.5368(10)&	17977	&	0.0408	&	\citet{Paschke+2021}\\
4300.4828(1)	&	4781	&	0.0000	&	\citet{2009IBVS.5898....1P}&9328.4664(7)&18133.5&0.0398&\citet{Lienhard+2022}	\\
4500.6209(3)	&	5312.5	&	-0.0006	&	\citet{2009IBVS.5898....1P}&9605.99382(1)&18870.5&0.0464&\citet{2022OEJV..234....1N}\\
4504.5751(8)	&	5323	&	-0.0002	&	\citet{2009IBVS.5898....1P}&10064.4578(30)&20088&0.0553&\citet{Paschke+2023}
\\
14513.6131(6)	&	5347	&	0.0005	&	\citet{2009IBVS.5898....1P}&\\
14560.1297	&	5470.5	&	0.0126	&	\citet{Nagai+2009}&	\\

\hline
\hline
\end{tabular}
\end{center}
\label{TabA}
\end{table*}


\end{document}